\documentclass[11pt]{article}

\usepackage[margin=1in]{geometry}
\usepackage[utf8]{inputenc}
\usepackage[T1]{fontenc}
\usepackage{lmodern}
\usepackage{microtype}
\usepackage{graphicx}
\usepackage{booktabs}
\usepackage{makecell}
\usepackage{amsmath}
\usepackage{xcolor}
\usepackage{listings}
\usepackage{caption}
\usepackage{subcaption}
\usepackage{float}
\usepackage{tikz}
\usetikzlibrary{arrows.meta,calc,fit,positioning}
\usepackage{enumitem}
\usepackage[numbers,sort&compress]{natbib}
\usepackage[colorlinks=true,allcolors=blue!55!black]{hyperref}
\usepackage{url}

\graphicspath{{figures/}}

\definecolor{codebg}{RGB}{247,249,251}
\definecolor{codekw}{RGB}{0,92,197}
\definecolor{codestr}{RGB}{3,120,80}
\definecolor{codecom}{RGB}{110,120,130}
\title{\bfseries AcadGIS: A Single-Import Python Package for Reproducible, Publication-Ready Academic Maps}
\author{%
  Ripon Chandra Malo\textsuperscript{1}, Shatabdi Roy\textsuperscript{2}, and Tong Qiu\textsuperscript{1}\\
  \small \textsuperscript{1} University of Utah, USA\\
  \small \textsuperscript{2} Jagannath University, Bangladesh
}
\date{July 2026}

\begin{document}
\maketitle

\begin{abstract}
Academic and project maps are often produced through a fragmented workflow: researchers locate boundaries, manage shapefiles, join tabular data, assemble locator insets, add cartographic decorations, and export figures through desktop GIS or multi-package Python scripts. This creates an accessibility barrier for non-GIS users and a reproducibility problem when data sources, styling choices, and manual edits are not captured in executable form. We present AcadGIS, a free and open-source Python package that creates publication-oriented research maps from high-level commands under one namespace, \texttt{import acadgis as agis}. AcadGIS provides place-name boundary access, automated study-area locator layouts, thematic cartography, raster and vector layers, curated Earth-observation products, terrain and hydrology context, and configurable PNG, PDF, and SVG export without requiring desktop GIS expertise or hand-managed shapefiles. Its design combines one-import access to the scientific-Python stack, publication-oriented defaults with progressive control, local caching, source attribution, and figure specifications based on code, named data, and a pinned package version. Through three representative use cases, we demonstrate how common paper, thesis, and project maps can be expressed as compact, inspectable scripts. Source code: \url{https://github.com/riponcm/AcadGIS}.
\end{abstract}

\vspace{0.5em}
\vspace{1em}

\section{Introduction}
\label{sec:introduction}

Maps are common components of scientific argument. A study-area figure tells the reader where fieldwork or project activity occurs; a choropleth turns a table into spatial evidence; terrain, land-cover, satellite, river, and road layers place local observations in geographic context. These figures appear throughout the earth, environmental, health, agricultural, engineering, and social sciences. For many students and researchers, however, producing a clear study-area or research map is a practical obstacle rather than a scientific task. They may know the place name and the data they want to show, but not how to obtain administrative boundaries, manage shapefiles, use desktop GIS, choose cartographic decorations, or combine multiple Python geospatial packages into a figure suitable for publication.

The difficulty is not a lack of map-making tools. Web map services can generate a quick location map, and desktop GIS systems such as QGIS \citep{qgis2024} and ArcGIS Pro \citep{arcgispro} provide mature cartographic environments. These tools are powerful, but they are not always aligned with the needs of a non-GIS specialist who simply needs a reproducible study map for a paper, thesis, project report, or proposal. A publication map commonly requires administrative boundaries from one source, imagery or elevation from another, tabular measurements from a spreadsheet, and final composition through a graphical interface or custom code layered over Matplotlib \citep{hunter2007matplotlib}, GeoPandas \citep{jordahl2020geopandas}, Rasterio, Contextily, PySAL, and other libraries. Each component is useful, but the combined workflow requires GIS knowledge, shapefile handling, package coordination, and manual styling decisions.

This creates two related gaps. The first is an accessibility gap: many researchers need publication-oriented maps but do not need, and may not have time to learn, a full GIS workflow. The second is a reproducibility gap: when a figure is assembled through downloads, point-and-click editing, and one-off scripts, the finished image often preserves the visual result but not the exact recipe. In an audit of published maps, \citet{koukouraki2023map} found that only $4$ of $27$ maps (roughly $15\%$) could be regenerated from the available materials. Maps encode data selections, administrative boundary versions, projections, classification schemes, scene dates, labels, and cartographic styling choices that shape how a result is interpreted. Reconstructing those choices also runs against established reproducible-research practice, which asks authors to record every step from raw input to final result and to prefer scripted, version-controlled workflows over manual manipulation \citep{sandve2013ten,wilson2017good}.

The research question motivating AcadGIS is therefore practical: Can the recurring maps needed in academic and project work be produced from high-level Python commands, without desktop GIS or manual shapefile management, while still preserving an inspectable figure specification? Existing Python libraries provide strong building blocks for geometry, plotting, classification, rasters, and basemaps, but they do not by themselves provide the complete workflow targeted here. Among the systems reviewed in this paper, we are not aware of another Python package designed specifically to combine place-name boundary access, study-area locator layouts, thematic cartography, contextual raster/vector layers, terrain, hydrology, publication styling, and export behind one high-level interface for non-GIS users.

AcadGIS answers this question by packaging recurring publication-map tasks as high-level Python calls under one namespace, \texttt{import acadgis as agis}. A user can name a place, add optional tabular or raster context, select a map type or layout template, and export a publication-oriented figure. The goal is not to replace full GIS platforms for advanced spatial analysis, but to make the common research-map figures in papers, theses, and project reports easier to create, inspect, and rerun.

\paragraph{Contributions.} This paper contributes:
\begin{enumerate}[label=(\roman*)]
  \item a high-level Python workflow for publication-oriented research maps aimed at users who need study-area, thesis, paper, or project maps without adopting a full desktop-GIS workflow or managing shapefiles;
  \item place-name access to administrative boundaries and bundled offline examples, reducing the need for manual shapefile search and preprocessing;
  \item automated study-area and locator layouts that generate nested country-to-site figures with insets, connectors, highlights, and per-panel context;
  \item a unified catalog of static research-map types, including choropleths, graduated symbols, heatmaps, isopleths and contours, dot-density maps, bivariate maps, cartograms, terrain maps, satellite and land-cover maps, rivers, roads, and sampling-site overlays; and
  \item a reproducible map specification based on code, named data, pinned package versions, local caching, and explicit source attribution.
\end{enumerate}

\section{Related Work}
\label{sec:related}

Related work spans three kinds of systems: full GIS environments, Python geospatial building blocks, and high-level mapping libraries. These systems overlap with AcadGIS in important ways, but they optimize for different users and workflows. The central distinction is that AcadGIS targets the recurring static maps needed in papers, theses, and project reports, especially for Python users who need publication-quality output without becoming GIS specialists.

\paragraph{Desktop, web, and cloud GIS workflows.}
General-purpose web maps can quickly locate a place, but they are not designed to produce reproducible study-area figures with administrative hierarchies, thematic joins, scale bars, north arrows, cited data sources, and journal export. Desktop GIS applications such as QGIS \citep{qgis2024} and ArcGIS Pro \citep{arcgispro} provide mature cartographic environments and remain the most capable tools for interactive GIS work. Their strength is also their cost for many academic users: learning GIS concepts, finding and managing geospatial files, documenting manual styling decisions, and reproducing the exact sequence of graphical edits. Cloud platforms such as Google Earth Engine \citep{gorelick2017gee} address a different need, namely large-scale Earth-observation analysis; they are less focused on the small number of static maps a researcher must place in a paper or thesis.

\paragraph{Python geospatial primitives.}
The Python ecosystem provides excellent lower-level components. Matplotlib \citep{hunter2007matplotlib} renders figures; NumPy \citep{harris2020numpy} and pandas \citep{mckinney2010pandas} handle arrays and tables; GeoPandas \citep{jordahl2020geopandas} brings simple-feature geometries into the pandas data model through Shapely \citep{gillies2007shapely}; and Rasterio \citep{gillies2013rasterio} reads and writes raster data. Cartopy \citep{metoffice_cartopy} supports projections and geodesic transforms, Contextily \citep{arribasbel_contextily} adds web-tile basemaps, and the PySAL ecosystem, including mapclassify \citep{rey2023geographic,rey2010pysal}, supplies standard classification schemes for choropleths. These packages are powerful, scriptable, and scientifically important, but they are primitives rather than a research-map application. A non-GIS author must still decide where to obtain boundaries, how to join names, how to construct locator panels, how to add cartographic decorations, and how to combine raster, vector, and thematic layers into one publication figure.

\paragraph{High-level mapping libraries.}
Several higher-level libraries reduce this burden, but their main workflows differ from AcadGIS. Folium \citep{folium} produces Leaflet-based interactive HTML maps for exploration and web delivery rather than print-first figures. PyGMT provides a Python interface to the Generic Mapping Tools and is particularly strong for static scientific and geophysical figures \citep{pygmt}. Geemap and leafmap reduce the code needed for interactive notebook mapping and Earth-observation analysis \citep{wu2020geemap,wu2021leafmap}. The R ecosystem provides a close analogue: sf \citep{pebesma2018sf} supplies the simple-features data model, and tmap \citep{tennekes2018tmap} offers a coherent grammar for thematic mapping with sensible cartographic defaults. These systems demonstrate the value of high-level mapping interfaces, but none centers on the particular combination of named administrative drill-down layouts, static thematic maps, and curated contextual layers targeted by AcadGIS.

\paragraph{Reproducibility and publication maps.}
The need for such a workflow is also a reproducibility issue. \citet{koukouraki2023map} found that only 4 of 27 examined published maps could be regenerated from available materials, reflecting the difficulty of reconstructing toolchains, data sources, classifications, and manual styling. General provenance systems for visualization, such as yProv4DV \citep{padovani2026yprov4dv}, can record execution context and inputs for Python visualization scripts. AcadGIS is complementary: instead of adding provenance around arbitrary plotting code, it narrows the task to academic cartography and makes common map figures expressible as named data, explicit options, cached sources, and a package version.

\paragraph{Positioning of AcadGIS.}
The gap is therefore not the absence of geospatial capability, but the absence, among the systems reviewed here, of a ready-made Python workflow focused on the recurring static maps that appear in academic and project documents. AcadGIS combines place-name boundary access, automated study-area and locator layouts, a thematic-map catalog grounded in cartographic theory \citep{slocum2009thematic} and color guidance \citep{harrower2003colorbrewer}, common terrain and Earth-observation layers, cartographic decorations, and publication export behind one import. It is built to occupy this specific space between desktop GIS, interactive notebook mapping, and lower-level scientific plotting libraries.

\section{Design and Workflow Architecture}
\label{sec:design}

AcadGIS is designed as a workflow layer over the Python geospatial stack. Its purpose is not to replace GeoPandas, Matplotlib, Rasterio, or desktop GIS for all spatial tasks, but to make the recurring research-map workflow explicit: a user names a place, optionally supplies a table or raster layer, chooses a map pattern, and exports a figure whose inputs and styling choices can be recovered from code. The design therefore emphasizes user-facing map specification rather than low-level geometry operations.

\subsection{Design Principles}

\paragraph{One import for the research-map workflow.}
The entire workflow is reached through one statement, \lstinline|import acadgis as agis|. The package also re-exports the scientific stack it builds on as attributes of the same namespace: \lstinline|agis.plt| (Matplotlib~\citep{hunter2007matplotlib}), \lstinline|agis.np| (NumPy), \lstinline|agis.pd| (pandas), and \lstinline|agis.gpd| (GeoPandas~\citep{jordahl2020geopandas}). A figure can therefore be assembled, annotated, joined to a table, and exported without a long import block. This matters pedagogically as well as technically: the map recipe remains short enough for a methods section, tutorial, or thesis appendix.

\paragraph{Ready-made patterns with progressive control.}
AcadGIS treats common academic maps as named patterns: a study-area locator, a choropleth, a terrain context map, a satellite or land-cover layer, a dot-density map, or a sampling-site map. Each pattern has publication-oriented defaults, but the same call can be refined through a consistent convention: booleans enable sensible defaults, named strings select curated variants, and dictionaries expose full control. For example, a north arrow can be turned on with \lstinline|True|, selected as \lstinline|"rose"|, or positioned and styled with a dictionary. Here, \emph{publication-oriented} means that figure size, resolution, labels, legends, standard cartographic decorations, palettes, and PNG/PDF/SVG export are controlled in code; compliance with a particular venue's dimensions and accessibility requirements remains the author's responsibility.

\paragraph{Reproducibility, local use, caching, and attribution.}
A figure is specified by code, named data, and the package version. Boundaries are requested by country and administrative level rather than by a hand-managed shapefile path. Bangladesh, Iraq, India, and the USA ship bundled, so their maps render with no network access; other countries are downloaded by name and cached on disk. Curated layers such as Natural Earth rivers, Copernicus DEM terrain, ESA WorldCover land cover, and Sentinel-2 NDVI are added through explicit calls. The documentation identifies these providers and their citation requirements so authors can report the sources used by each figure.

\subsection{Workflow Architecture}

\begin{figure}[t]
\centering
\resizebox{\linewidth}{!}{%
\begin{tikzpicture}[
  font=\footnotesize,
  >={Latex[length=2.1mm]},
  node distance=0.55cm and 0.65cm,
  stage/.style={draw=black!50, rounded corners=2pt, align=center, inner sep=5pt, minimum height=1.05cm, text width=3.05cm},
  input/.style={stage, fill=blue!6},
  service/.style={stage, fill=green!7},
  source/.style={stage, fill=orange!9},
  output/.style={stage, fill=purple!7},
  lab/.style={font=\bfseries\small, align=center}
]
\node[input] (spec) {Research map request\\place names\\table or raster\\map type and options};
\node[service, right=of spec] (resolve) {Resolve places\\load boundaries\\match table names\\cache sources};
\node[service, right=of resolve] (construct) {Build map pattern\\locator\\choropleth\\terrain and layers\\decorations};
\node[output, right=of construct] (figure) {Publication figure\\PNG, PDF, SVG\\legend\\scale bar\\north arrow};

\node[source, below=0.85cm of resolve] (data) {Curated data\\bundled countries\\GADM\\Natural Earth\\OSM, Copernicus\\ESA, Sentinel-2};
\node[source, below=0.85cm of construct] (stack) {Python stack\\GeoPandas, Matplotlib\\pandas, NumPy\\Rasterio, PySAL};
\node[output, below=0.85cm of figure] (record) {Reproducible record\\script\\named data\\package version\\source citations};

\node[lab, above=0.24cm of spec] {User specification};
\node[lab, above=0.24cm of resolve] {AcadGIS workflow};
\node[lab, above=0.24cm of figure] {Output};

\draw[->, thick] (spec.east) -- (resolve.west);
\draw[->, thick] (resolve.east) -- (construct.west);
\draw[->, thick] (construct.east) -- (figure.west);
\draw[->, thick] (figure.south) -- (record.north);
\draw[->, dashed] (data.north) -- (resolve.south);
\draw[->, dashed] (data.north east) -- (construct.south west);
\draw[->, dashed] (stack.north) -- (construct.south);
\draw[->, dashed] (stack.north east) -- (figure.south west);
\end{tikzpicture}%
}
\caption{AcadGIS workflow architecture. A researcher supplies a named map specification; AcadGIS resolves places, joins data, constructs a publication-oriented map pattern, and exports the figure. The corresponding script, named data, package version, and documented sources constitute its reproducibility record. Dashed arrows indicate supporting data and Python libraries used by the workflow.}
\label{fig:architecture}
\end{figure}
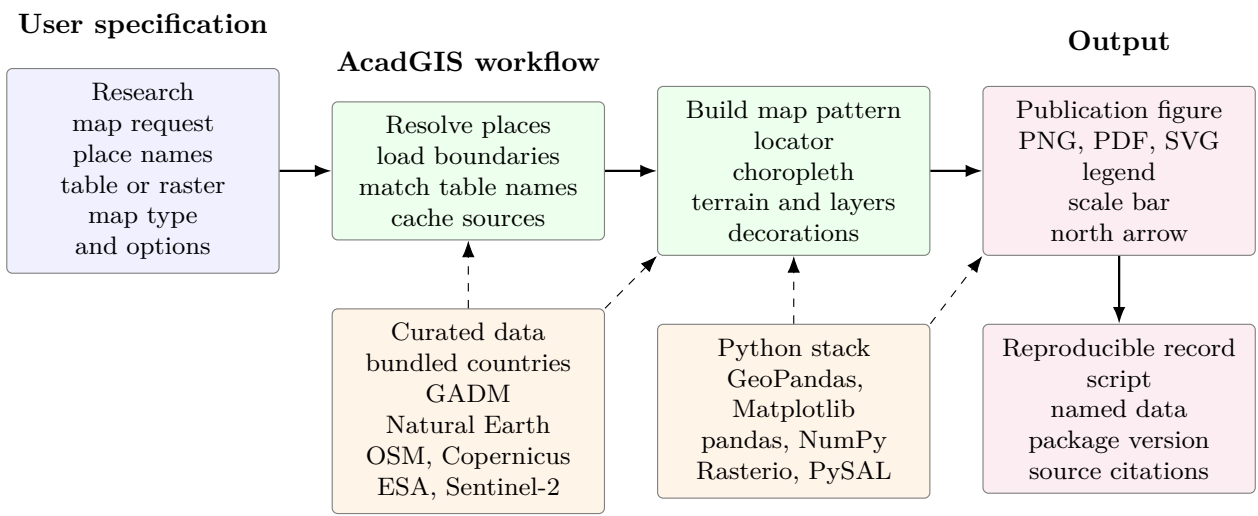

Figure~\ref{fig:architecture} summarizes the workflow architecture. The user-facing inputs are intentionally close to the language of a paper: a place name, optional measurements, a desired map type, and styling or export options. AcadGIS resolves those inputs through four internal services. First, the boundary and cache service obtains administrative geometries and avoids repeated downloads. Second, the matching service joins user tables to place names and reports unmatched records. Third, map constructors create locator layouts, thematic maps, contextual layers, terrain, and hydrology. Fourth, the styling and export service applies cartographic defaults and writes the final PNG, PDF, or SVG. The underlying scientific stack remains available throughout, so advanced users can keep editing the returned Matplotlib axes or GeoDataFrames.

Listing~\ref{lst:disclosure} illustrates the same idea in code: a simple map can be produced with defaults, and then refined by replacing a boolean decoration with a structured option dictionary without changing the surrounding workflow.

\begin{lstlisting}[language=Python,caption={Single import and progressive disclosure: booleans first, then a dict for fine control.},label={lst:disclosure}]
import acadgis as agis

gdf = agis.load_boundaries("Bangladesh", level="district")

# Defaults on with booleans
agis.plot(gdf, north_arrow=True, scale_bar=True, border=True)

# Same call, one decoration refined with a dict
agis.plot(gdf, north_arrow={"style": "compass", "loc": "upper right"},
          scale_bar=True, border=True)
\end{lstlisting}

\section{Functionality}
\label{sec:functionality}

AcadGIS functionality is organized around the sequence by which research maps are normally produced: a place is identified, boundaries and contextual data are resolved, the appropriate map pattern is selected, and the figure is styled and exported. This section therefore describes capabilities by workflow role rather than by module name. The same interface convention is used throughout: simple calls produce usable defaults, while advanced users can refine the result with named styles or option dictionaries.

\subsection{Administrative boundaries and fuzzy name matching}
\label{subsec:boundaries}

The entry point for most figures is \lstinline|load_boundaries(country, level, within=...)|, which returns administrative units as a GeoDataFrame. The user specifies a country and a familiar administrative level---for example country, state or division, district, or upazila---rather than a local shapefile path. The optional \lstinline|within| argument restricts the result to a named parent unit, allowing a script to request, for example, districts within Dhaka division. Bangladesh, Iraq, India, and the USA are bundled for local use; other countries are downloaded by name from GADM \citep{gadm} and cached, so later runs use the local copy. Physical reference layers such as coastlines are drawn from Natural Earth \citep{naturalearth}.

The same place-name orientation is used when research tables are joined to boundaries. This is an important practical step because administrative names change, vary by source, or appear with omitted suffixes and diacritics. Exact string joins can therefore produce a visually plausible but incomplete map. AcadGIS provides \lstinline|match_one| and \lstinline|match_table|, backed by \texttt{rapidfuzz}, and surfaces unmatched records rather than silently dropping them. The \lstinline|choropleth| helper wraps this matching, classification, and drawing step so that a spreadsheet can be mapped directly by place name (Listing~\ref{lst:choropleth}).

\begin{lstlisting}[language=Python,caption={A choropleth joined to US states by name, classified into quantiles.},label={lst:choropleth}]
import acadgis as agis

states = agis.load_boundaries("USA", level=1)
income = agis.pd.read_csv("income.csv")
ax = agis.choropleth(states, income, key="state",
                     value="median_income", scheme="quantiles",
                     palette="viridis", legend=True)
agis.save(ax, "income.pdf", dpi=300)
\end{lstlisting}

Classification schemes, including quantiles, natural breaks and equal interval classes, are supplied through \texttt{mapclassify} from the PySAL family \citep{rey2023geographic,rey2010pysal}. Curated palettes include colorblind-aware choices informed by ColorBrewer guidance \citep{harrower2003colorbrewer}; authors can override them and remain responsible for checking accessibility in the intended medium.

\subsection{Study-area locator layouts}
\label{subsec:locator}

The study-area locator is the figure type most directly motivated by academic writing. Papers, theses, and project reports often need to situate a site within a country, region, and local administrative unit, but the resulting composition normally requires insets, repeated extents, connector boxes, labels, scale bars, and consistent styling. AcadGIS provides two interfaces for this task. The fluent \lstinline|StudyArea| builder supports a drill-down expression such as \lstinline|StudyArea("India", context_level="state").zoom_into("West Bengal", detail_level="district").figure()|. The one-call \lstinline|study_area| interface produces the same family of figures from a list of named administrative steps (Listing~\ref{lst:studyarea}). Five templates cover common publication compositions: \texttt{single}, \texttt{two}, \texttt{cascade}, \texttt{series}, and \texttt{grid}.

\begin{lstlisting}[language=Python,caption={A series locator from country to district using named administrative steps.},label={lst:studyarea}]
import acadgis as agis

fig = agis.study_area("Bangladesh",
        steps=[("division", "Dhaka"), ("district", "Gazipur")],
        template="series",
        rivers=True, labels=True)
agis.save(fig, "studyarea.png", dpi=300)
\end{lstlisting}

\begin{figure}[H]
\centering
\includegraphics[width=\linewidth]{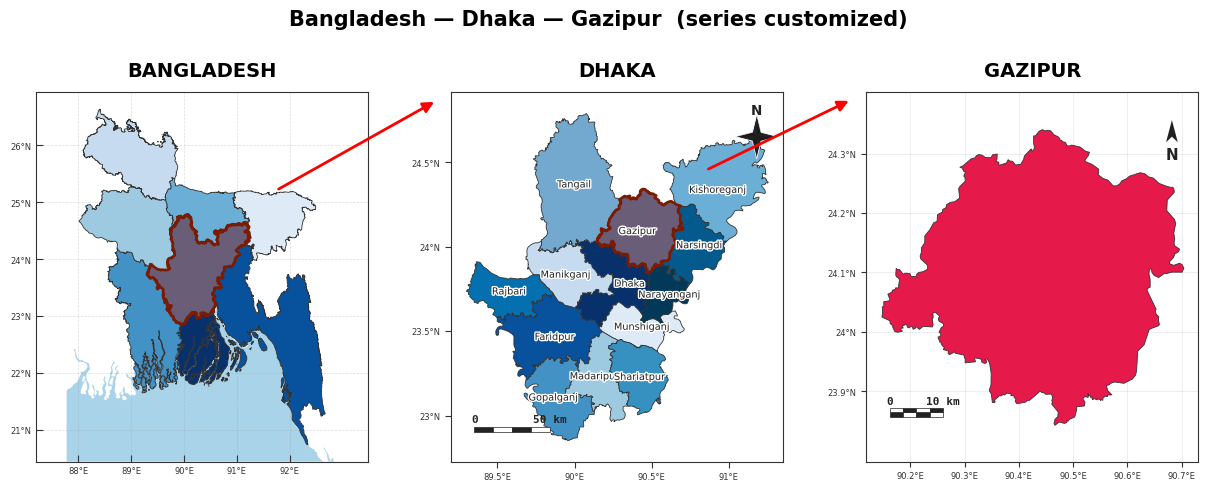}
\caption{Series study-area locator for Bangladesh\,$\rightarrow$\,Dhaka Division\,$\rightarrow$\,Gazipur District. Three equal panels are linked by connector arrows, and each includes a north arrow and scale bar (Listing~\ref{lst:studyarea}).}
\label{fig:studyarea}
\end{figure}

AcadGIS resolves each named unit, nests the locator panels (country\,$\rightarrow$\,region\,$\rightarrow$\,site), and draws the connector boxes and arrows that tie each context view to the next detail view. Connectors can be enlarged, shrunk, stretched, anchored by figure corner or geographic coordinate, and styled with or without arrowheads and endpoint markers. Sea, rivers, labels, north arrows, and scale bars can be enabled globally or per panel. Figure~\ref{fig:studyarea} shows a series layout from Bangladesh through Dhaka division to Gazipur district; the same named-step convention supports the other locator templates when their required panel structure is used.

\subsection{Thematic map catalog}
\label{subsec:thematic}

\begin{figure}[H]
\centering
\begin{subfigure}[t]{0.49\linewidth}
\centering
\includegraphics[width=\linewidth]{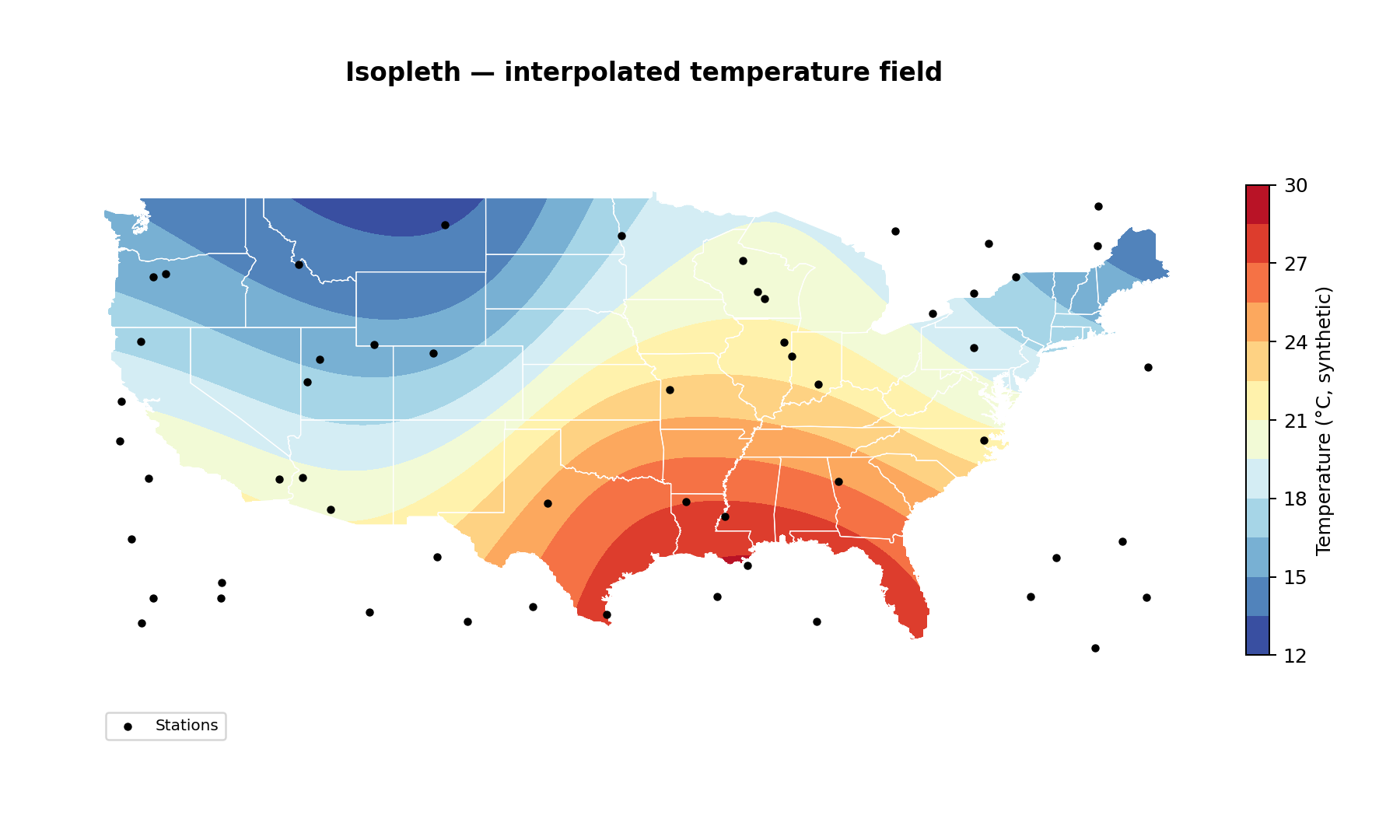}
\caption{}
\label{fig:thematic-isopleth}
\end{subfigure}
\hfill
\begin{subfigure}[t]{0.49\linewidth}
\centering
\includegraphics[width=\linewidth]{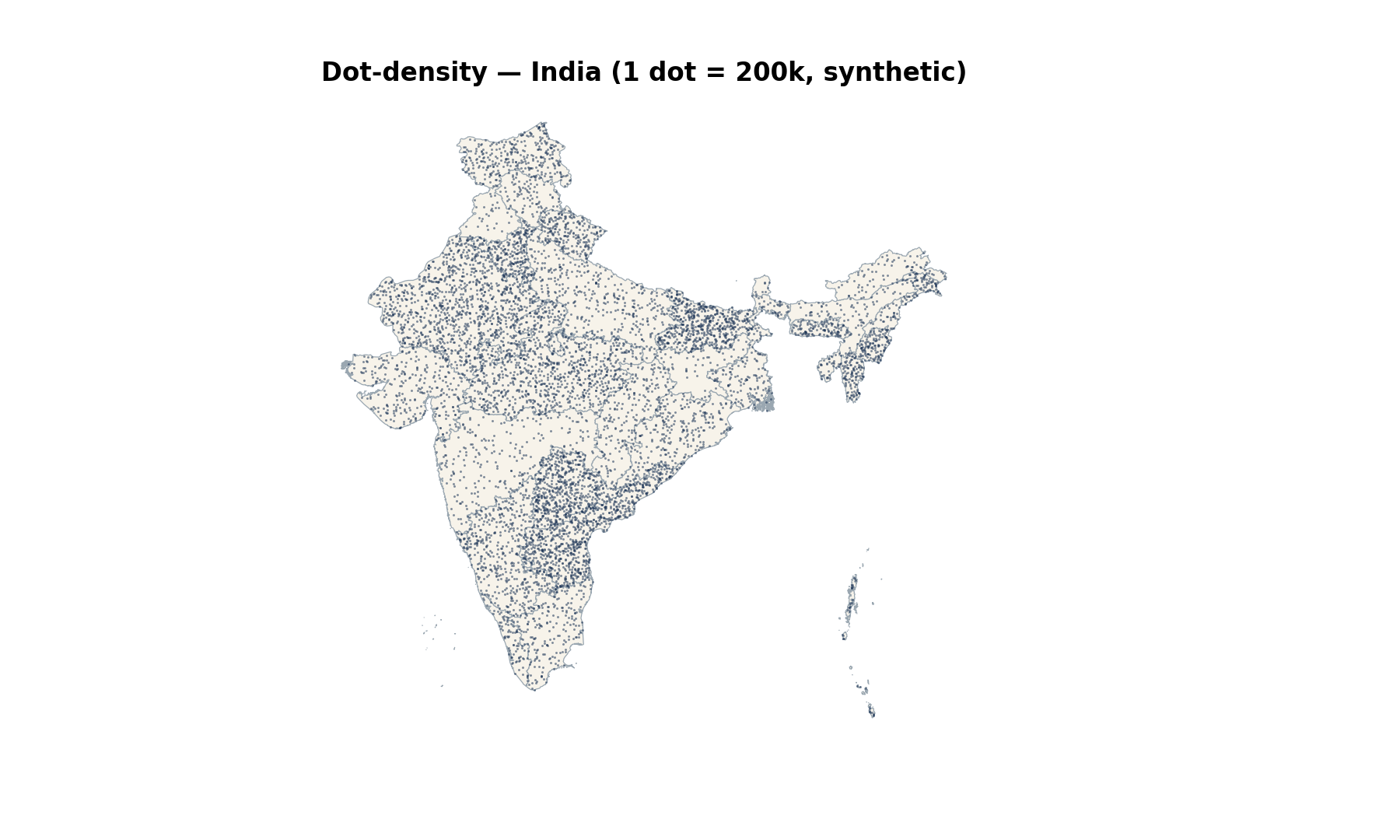}
\caption{}
\label{fig:thematic-dot-density}
\end{subfigure}

\vspace{2pt}
\begin{subfigure}[t]{0.49\linewidth}
\centering
\includegraphics[width=\linewidth]{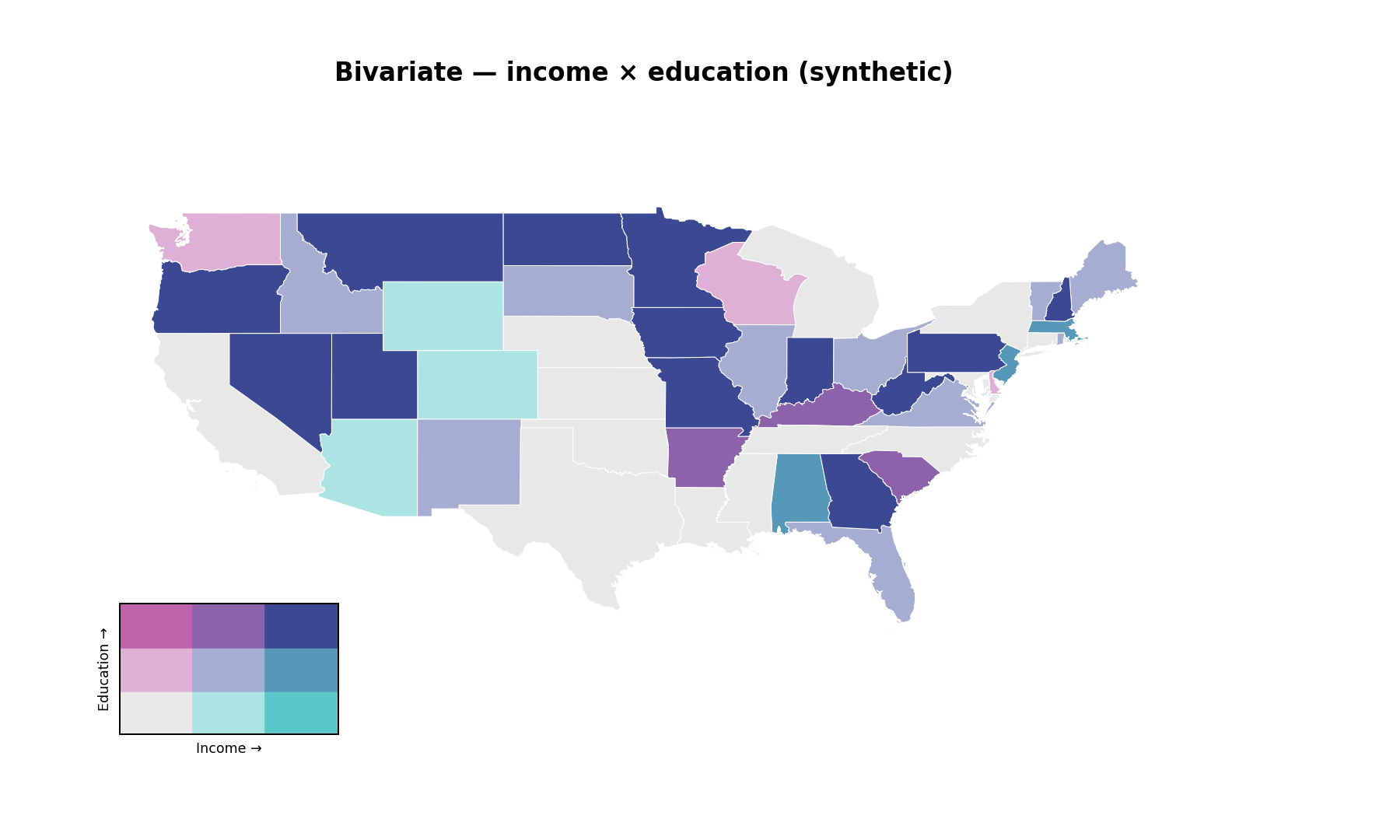}
\caption{}
\label{fig:thematic-bivariate}
\end{subfigure}
\hfill
\begin{subfigure}[t]{0.49\linewidth}
\centering
\includegraphics[width=\linewidth]{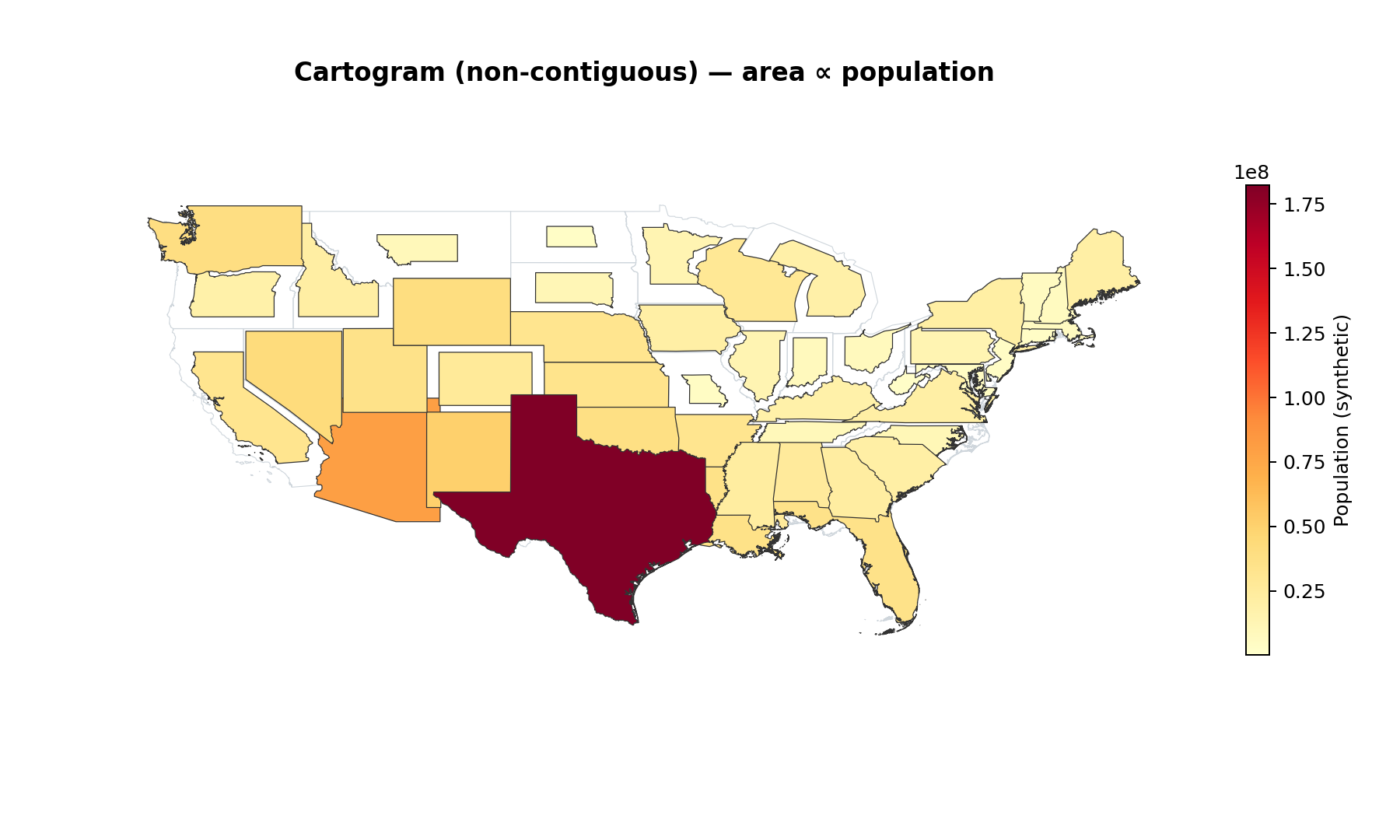}
\caption{}
\label{fig:thematic-cartogram}
\end{subfigure}
\caption{Examples from the thematic-map catalog: (a) an interpolated temperature isopleth; (b) a dot-density map of India, with one dot representing 200{,}000 units; (c) a bivariate choropleth of income and education; and (d) a non-contiguous cartogram scaled by population. All thematic values are synthetically generated to demonstrate the map types and do not represent observed measurements.}
\label{fig:thematic}
\end{figure}

Beyond the choropleth, AcadGIS implements a catalog of standard static thematic maps \citep{slocum2009thematic} as scripted operations on a common plotting surface. Density can be shown with \lstinline|add_heatmap| as a kernel-density surface or hexbin. Continuous fields can be estimated from scattered observations with \lstinline|interpolate_field| using NumPy-based Gaussian-kernel interpolation \citep{harris2020numpy}, then rendered by \lstinline|add_isopleth| or \lstinline|add_contours|. Count data can be represented with \lstinline|dot_density| at an explicit dot-to-count ratio. Two-variable spatial patterns can be shown with \lstinline|bivariate| using a $3\times3$ color matrix and matching legend, while \lstinline|cartogram| produces Dorling or non-contiguous cartograms scaled by an attribute. Graduated proportional symbols are available through \lstinline|points|. Figure~\ref{fig:thematic} illustrates four methods that would otherwise require separate recipes but are exposed through the same AcadGIS plotting convention.

\subsection{Raster and vector layers and curated Earth-observation data}
\label{subsec:layers}

Research maps often need more than one data source: a boundary layer, a satellite backdrop, a road or river network, and a raster product may all appear in the same figure. AcadGIS treats these as executable layers added to an existing axis, so contextual information remains part of the script rather than a later graphical edit. \lstinline|add_raster| draws GeoTIFF data---continuous, categorical, or RGB---and reprojects it on the fly through \texttt{rioxarray} and \texttt{rasterio} \citep{gillies2013rasterio}. \lstinline|add_layer| adds arbitrary vector data while detecting polygon, line, or point geometry. Tile basemaps are exposed through \lstinline|add_basemap| and \lstinline|add_satellite| using \texttt{contextily} and \texttt{xyzservices} \citep{arribasbel_contextily}. Cities, roads, and other reference features can be added through \lstinline|add_cities|, \lstinline|load_places|, and \lstinline|add_roads|.

\begin{figure}[H]
\centering
\includegraphics[height=0.62\textheight,keepaspectratio]{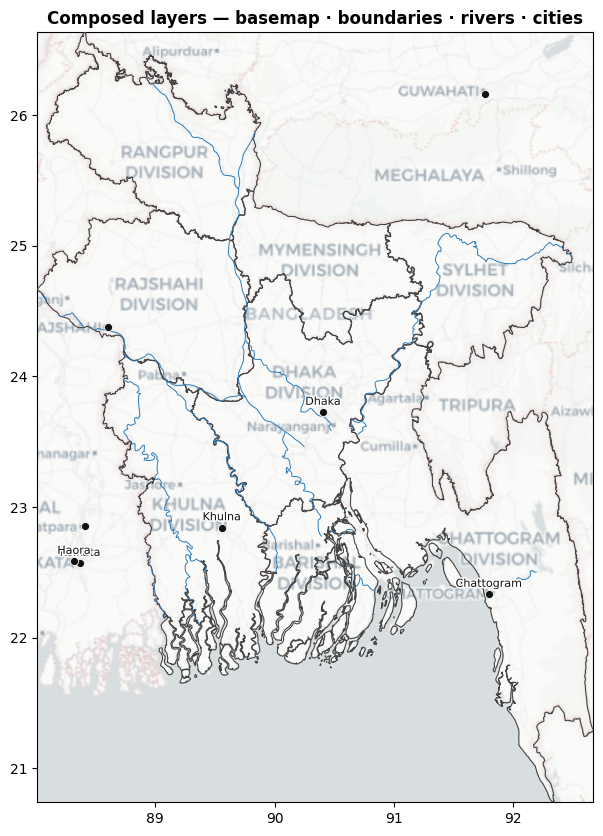}
\caption{A composed map: a satellite tile basemap overlaid with administrative boundaries, rivers and labelled cities, each added with a single call.}
\label{fig:layers}
\end{figure}

Curated Earth-observation layers are folded into the same workflow. \lstinline|add_landcover| reads ESA WorldCover at 10\,m resolution \citep{zanaga2022worldcover} using windowed cloud-optimized GeoTIFF access, so only the mapped extent is fetched. \lstinline|add_ndvi| retrieves Sentinel-2 imagery \citep{drusch2012sentinel2} through the Earth Search STAC catalog and computes the vegetation index in memory. Figure~\ref{fig:layers} shows a map composed from a satellite basemap, boundaries, rivers, and cities.

\subsection{Terrain and hydrology}
\label{subsec:terrain}

\begin{figure}[H]
\centering
\includegraphics[width=0.7\linewidth]{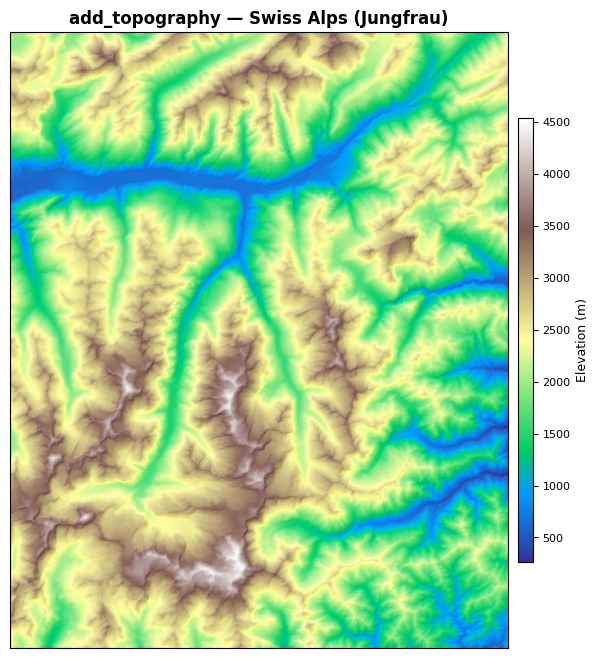}
\caption{Shaded relief with hypsometric tint over the Swiss Alps, rendered by \lstinline|relief| from the Copernicus GLO-30 DEM.}
\label{fig:terrain}
\end{figure}

Terrain support provides physical context without requiring a separate DEM-processing workflow. \lstinline|load_dem| fetches Copernicus GLO-30 elevation tiles for an extent \citep{copernicusdem}, and \lstinline|relief| renders shaded relief with hypsometric tint and land-to-ocean coloring. Hydrographic context can be added with \lstinline|add_rivers| from Natural Earth \citep{naturalearth} or OpenStreetMap \citep{osm}, together with \lstinline|add_sea| and \lstinline|add_water|. For studies where drainage is part of the analysis rather than only background context, \lstinline|drainage| and \lstinline|add_streams| derive and draw stream networks from DEM flow accumulation using \texttt{pysheds}. Figure~\ref{fig:terrain} shows shaded relief of the Swiss Alps.

\subsection{Styling and export}
\label{subsec:export}

All maps render on Matplotlib \citep{hunter2007matplotlib} and return standard figure or axis objects, so users can continue editing with Matplotlib through the re-exported \lstinline|agis.plt| when needed. The built-in styling system provides palettes, themes, north-arrow styles, scale-bar styles, borders, graticules, legends, and labels. Each can be enabled with a boolean, selected by a short name, or customized with an option dictionary. Because the map is specified by code, named data, and the package version, the same script can regenerate the figure and expose the choices behind it. Export is handled by \lstinline|agis.save(obj, path, dpi=300)|, which writes PNG, PDF, or SVG output from an AcadGIS figure or Matplotlib object.

\section{Use-Case Demonstrations}
\label{sec:demonstrations}

We demonstrate AcadGIS through three representative mapping tasks rather than presenting runtime, usability, or comparative benchmarks. The examples show how common academic map requirements can be expressed as short, inspectable scripts: a study-area locator that would normally require manual layout, a thematic map joined from a spreadsheet by name, and a remote-sensing context map that would otherwise require scene search and band processing. Each task is presented with executable code and the resulting map. The script, named data sources, package version ($0.2.0$), and documented upstream providers form the reproducibility record.

\begin{table}[H]
\centering
\caption{Representative use cases used to demonstrate the AcadGIS workflow.}
\label{tab:evaluation-tasks}
\small
\begin{tabular}{@{}>{\raggedright\arraybackslash}p{0.24\linewidth}>{\raggedright\arraybackslash}p{0.29\linewidth}>{\raggedright\arraybackslash}p{0.35\linewidth}@{}}
\toprule
Task & Common manual burden & Demonstrated output \\
\midrule
Study-area locator & Selecting boundaries, arranging panels, drawing connectors, adding map decorations & Listing~\ref{lst:cs-studyarea} and Figure~\ref{fig:grid_drilldown} generate a country-to-site layout from named units. \\
Spreadsheet choropleth & Matching place names, classifying values, selecting a palette, exporting a figure & Listing~\ref{lst:cs-choropleth} and Figure~\ref{fig:choropleth_income} map tabular values with fuzzy matching and named classification. \\
Sentinel-2 NDVI map & Searching scenes, downloading bands, computing an index, recording scene choices & Listing~\ref{lst:ndvi} and Figure~\ref{fig:ndvi_nile} compute NDVI from a bounding box and date window. \\
\bottomrule
\end{tabular}
\end{table}

\subsection{Use Case 1: Study-Area Locator From Named Units}

Many field-based studies open with a figure that locates the study site within a wider administrative hierarchy. Producing this by hand involves nesting insets, drawing connector boxes, choosing panel extents, and keeping projections consistent across panels, all of which are easy to lose from the reproducibility record \citep{koukouraki2023map}. Listing~\ref{lst:cs-studyarea} drills from Bangladesh through Dhaka division and Gazipur district to Sreepur upazila; \texttt{study\_area} resolves each name against GADM boundaries \citep{gadm}, chooses matching extents, and draws the connectors, north arrow, scale bar, graticule, and river context.

\begin{lstlisting}[language=Python,caption={A four-level study-area drill-down produced by one call.},label={lst:cs-studyarea}]
import acadgis as agis

fig = agis.study_area(
    "Bangladesh",
    steps=[("division", "Dhaka"),
           ("district", "Gazipur"),
           ("upazila",  "Sreepur")],
    template="grid", labels=True, rivers=True)
agis.save(fig, "grid_drilldown.png", dpi=300)
\end{lstlisting}

\begin{figure}[H]
\centering
\includegraphics[width=\linewidth]{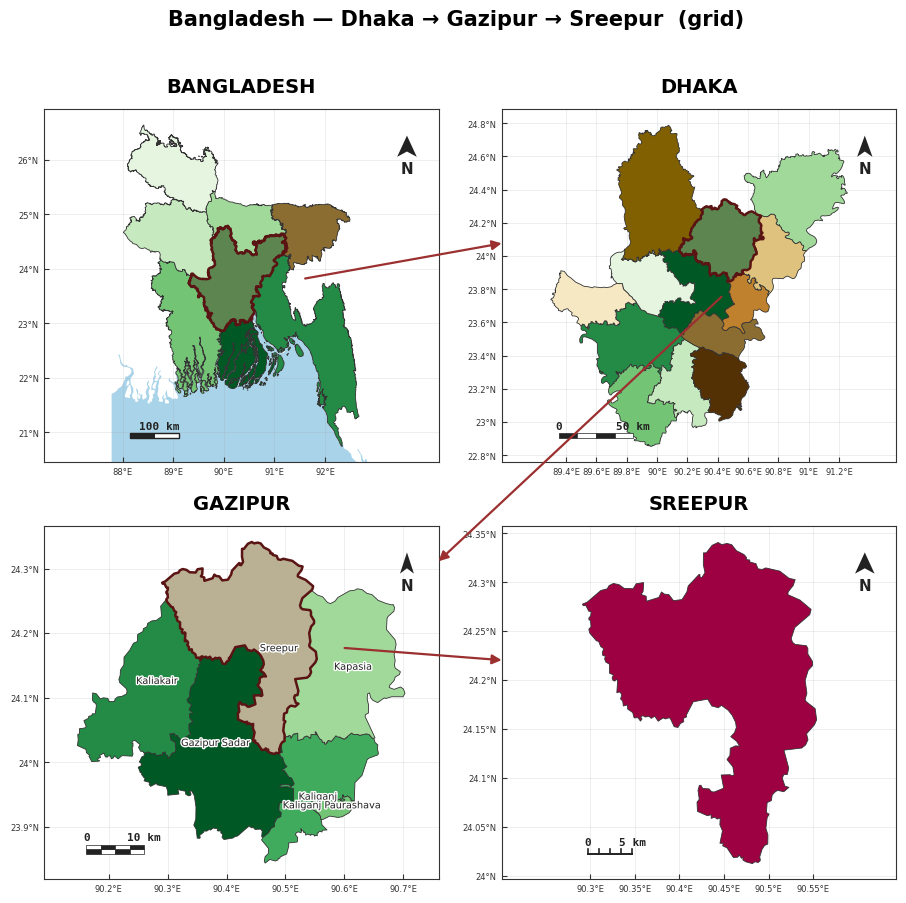}
\caption{Study-area cascade from Bangladesh to Sreepur upazila (\texttt{grid} template) via Dhaka division and Gazipur district. Each panel highlights its target unit; connectors, north arrow, scale bar, graticule, and river overlays are placed automatically by \texttt{study\_area} in Listing~\ref{lst:cs-studyarea}.}
\label{fig:grid_drilldown}
\end{figure}

The result (Figure~\ref{fig:grid_drilldown}) is a four-panel grid in which each panel highlights the target unit and connects it to the coarser panel. The code specifies no shapefile paths, coordinates, or inset geometry. The panels follow from the named units and template choice, so the same listing regenerates the layout wherever the boundaries are available or cached.

\subsection{Use Case 2: Choropleth Joined From a Spreadsheet}

Thematic values in the social and health sciences often arrive as a spreadsheet keyed by place name, not by a boundary identifier. Names may include diacritics, dropped administrative suffixes, spelling variants, or official renamings, so an exact string join can silently discard rows. AcadGIS performs a fuzzy name join inside \texttt{choropleth}, reports the units it could not match, and classifies the values with a named \texttt{mapclassify} scheme \citep{rey2023geographic} and a curated palette \citep{harrower2003colorbrewer}. Listing~\ref{lst:cs-choropleth} maps synthetic per-state values across the United States using natural (Fisher--Jenks) breaks.

\begin{lstlisting}[language=Python,caption={A choropleth joined by name from a DataFrame, with a classification scheme.},label={lst:cs-choropleth}]
import acadgis as agis

states = agis.load_boundaries("USA", level=1)
df = agis.pd.read_csv("income.csv")   # columns: state, income
ax = agis.choropleth(states, df, key="state", value="income",
                     scheme="fisher_jenks", k=5,
                     palette="viridis", legend=True)
agis.save(ax, "choropleth_income.png", dpi=300)
\end{lstlisting}

\begin{figure}[H]
\centering
\includegraphics[width=\linewidth]{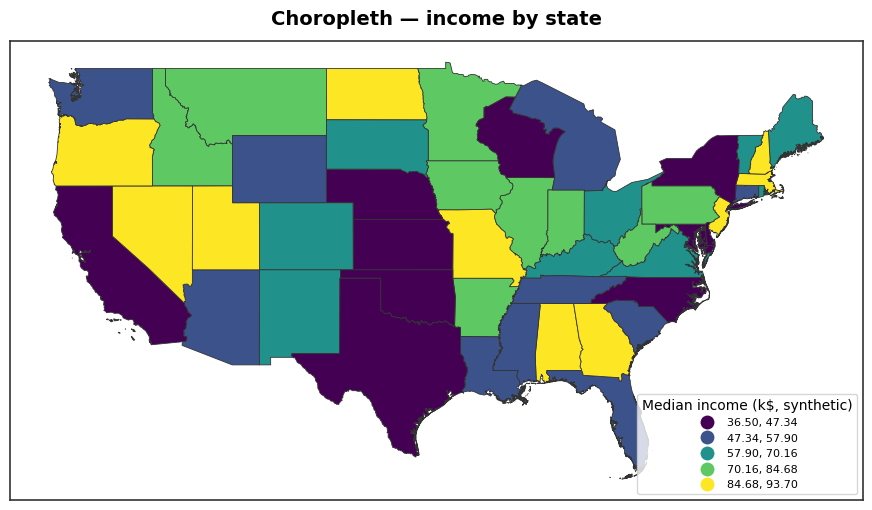}
\caption{Choropleth of synthetic per-state values for the United States, joined from a spreadsheet by state name with fuzzy matching and classified into five Fisher--Jenks classes (Listing~\ref{lst:cs-choropleth}). The values demonstrate the workflow and do not represent observed state income.}
\label{fig:choropleth_income}
\end{figure}

The spreadsheet needs no geometry column and no manual key alignment; unmatched names are surfaced rather than dropped without notice. Figure~\ref{fig:choropleth_income} shows the classified surface with its legend. Because the join key, classification scheme, class count, and palette are named arguments rather than interactive choices, the map can be reproduced from the listing and the two named inputs.

\subsection{Use Case 3: Sentinel-2 NDVI Computed On the Fly}

Vegetation and land-surface studies often need a normalized difference vegetation index (NDVI) for a small study area, but obtaining one can involve account registration, scene search, band download, band selection, and manual index calculation. \texttt{add\_ndvi} collapses this to one call over a bounding box: it queries the Earth Search STAC catalog for the least-cloudy Sentinel-2 scene \citep{drusch2012sentinel2} in a date window, reads the red and near-infrared bands with windowed access, and computes NDVI in memory, with no API key. Listing~\ref{lst:ndvi} maps the Nile at Luxor.

\begin{lstlisting}[language=Python,caption={On-the-fly Sentinel-2 NDVI for a bounding box, no API key.},label={lst:ndvi}]
import acadgis as agis

luxor = (32.55, 25.60, 32.95, 25.95)
fig, ax = agis.plt.subplots(figsize=(8.5, 8))
ax.set_xlim(luxor[0], luxor[2]); ax.set_ylim(luxor[1], luxor[3])
agis.add_ndvi(ax, luxor, start="2023-01-01", end="2024-12-31",
              max_cloud=5, cmap="RdYlGn")
agis.save(fig, "ndvi_nile.png", dpi=300)
\end{lstlisting}

\begin{figure}[H]
\centering
\includegraphics[height=0.42\textheight,keepaspectratio]{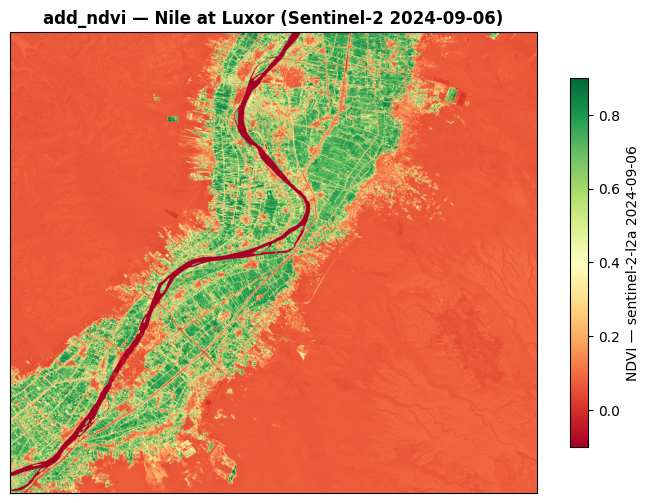}
\caption{Sentinel-2 NDVI along the Nile at Luxor, fetched from the Earth Search STAC catalog and computed on the fly by \texttt{add\_ndvi} (Listing~\ref{lst:ndvi}); irrigated fields form a high-index ribbon through the surrounding desert.}
\label{fig:ndvi_nile}
\end{figure}

Figure~\ref{fig:ndvi_nile} renders the irrigated Nile valley as a high-NDVI corridor through low-index desert. The chosen scene date is recorded on the axes so it can be reported with the figure, keeping the map traceable to a specific acquisition. The bounding box, cloud threshold, date window, and color map determine which scene is fetched and how the index is drawn.

\clearpage
\section{Discussion}
\label{sec:discussion}

The use-case demonstrations present AcadGIS as a workflow layer for recurring academic map figures, not as a replacement for full GIS environments or lower-level geospatial libraries. They establish functional examples, not evidence of runtime, usability, or cartographic superiority. This section therefore discusses where AcadGIS sits relative to existing tools, what remains outside its scope, and how the package can be extended.

\subsection{Position Relative to Existing Tools}

AcadGIS occupies a niche that existing tools address only partially. Table~\ref{tab:comparison} contrasts it with the building blocks and applications researchers commonly use when producing maps. The comparison is organized by workflow capability rather than by feature count, because the central claim is not that each cartographic primitive is new. The claim is that recurring academic map figures can be expressed through a coherent Python interface instead of a workflow scattered across boundary downloads, graphical layout, and bespoke plotting code.

\begin{table}[H]
\centering
\caption{Workflow-oriented comparison of AcadGIS with common map-making tools. Yes denotes a dedicated built-in workflow, partial denotes a capability achievable through manual configuration, scripting, plugins, or multiple operations, and no denotes that the capability is not a standard workflow of the tool. The categories describe typical use rather than every possible extension.}
\label{tab:comparison}
\small
\begin{tabular}{@{}lcccccc@{}}
\toprule
Capability & AcadGIS & \makecell{geopandas\\+mpl} & cartopy & folium & QGIS & \makecell{tmap\\(R)} \\
\midrule
Print-oriented defaults           & yes     & no      & partial & no      & partial & yes \\
Automated study-area/locator      & yes     & no      & no      & no      & partial & no \\
Multiple thematic map types       & yes     & partial & no      & partial & yes     & yes \\
Curated EO helper calls           & yes     & no      & no      & no      & partial & no \\
Terrain from DEM                  & yes     & no      & partial & no      & yes     & no \\
Reproducible / scriptable         & yes     & yes     & yes     & yes     & partial & yes \\
Interactive                       & no      & no      & no      & yes     & yes     & partial \\
Primarily for print figures       & yes     & yes     & yes     & no      & no      & yes \\
\bottomrule
\end{tabular}
\end{table}

The general-purpose Python stack pairs \texttt{geopandas}~\citep{jordahl2020geopandas} with Matplotlib and \texttt{cartopy}~\citep{metoffice_cartopy}. It is fully scriptable but exposes primitives rather than ready-made research-map workflows. A locator inset, a natural-breaks choropleth with an accessible palette, or a shaded-relief backdrop must each be assembled from those components. \texttt{folium}~\citep{folium} targets interactive HTML slippy maps and is therefore less suitable for static print figures. QGIS~\citep{qgis2024} supports reproducible projects, processing models, and Python automation, although manual layout operations require deliberate documentation. The closest analogue in spirit is \texttt{tmap}~\citep{tennekes2018tmap}, which offers a layered grammar with publication-quality output in R; however, it does not center on the same nested country$\rightarrow$region$\rightarrow$site locator workflow or curated Earth-observation helpers. AcadGIS is distinguished by combining these workflows within one Python namespace tuned for static academic figures.

\subsection{Scope and Limitations}

AcadGIS is deliberately scoped. It renders on a static Matplotlib backend~\citep{hunter2007matplotlib} and is aimed at publication figures rather than heavy interactivity; users needing pan-and-zoom web maps are better served by tools such as \texttt{folium}~\citep{folium}. It is also a cartographic package, not a spatial-analysis library: it does not implement spatial autocorrelation, regression, routing, or point-pattern analysis, deferring such work to specialized tools such as PySAL~\citep{rey2023geographic}. Projection support is curated for common academic use cases rather than exhaustive, so specialized coordinate systems may require dropping down to the underlying libraries. User-selected palettes and venue-specific output settings still require accessibility and formatting checks. Finally, while Bangladesh, Iraq, India, and the USA ship bundled for offline use, boundaries, tiles, and Earth-observation data for other regions are downloaded on demand and cached, so a network connection is required the first time a non-bundled area is mapped. Exact regeneration of remotely retrieved products also requires recording the provider, product version or item identifier, and acquisition date.

\subsection{Future Work}

Several directions extend the current release. The most visible is a no-code web application that exposes the same defaults and locator automation through a browser while preserving a scriptable specification behind each figure. On the cartographic side, planned work includes contiguous cartograms to complement the existing Dorling and non-contiguous variants, broader projection support, and a larger collection of bundled countries so more common study areas are available offline. A further direction is to make provenance reporting more explicit by exporting the selected data sources, package version, and figure options as a small sidecar record alongside each map.

The high-level AcadGIS API also creates an opportunity for AI-assisted map authoring. Future work will evaluate whether coding assistants can reliably translate natural-language map specifications into valid AcadGIS scripts and will support that interaction through machine-readable API schemas, validated prompt-to-map examples, parameter checking, and provenance records linking generated scripts to their data sources and package version.

\section{Conclusion}
\label{sec:conclusion}

AcadGIS addresses a practical gap between desktop GIS software and low-level Python geospatial stacks. Many researchers need study-area maps, thematic figures, terrain context, or satellite-derived products for papers and projects, but seek to avoid manually coordinating shapefiles, projections, symbolization, and layout. AcadGIS brings these recurring tasks into one Python namespace with publication-oriented defaults and progressive control. It automates country$\rightarrow$region$\rightarrow$site locator figures, provides multiple thematic map types, retrieves selected Earth-observation and terrain data through helper calls, and documents the underlying sources. The contribution is therefore not another map renderer alone, but an inspectable figure workflow: a short script, named data inputs, a package version, and pinned upstream products provide the basis for regenerating the map and reviewing the choices behind it. This makes academic maps easier to prepare, review, rerun, and share alongside the scientific result they communicate.

\section*{Data and Code Availability}
AcadGIS~0.2.0 is open-source under the Apache-2.0 license. Installable packages and source distributions are available on \href{https://pypi.org/project/acadgis/}{PyPI} (\texttt{pip install acadgis}), and the exact source is preserved in the \href{https://github.com/riponcm/AcadGIS/releases/tag/v0.2.0}{GitHub v0.2.0 release}. Development source is available in the \href{https://github.com/riponcm/AcadGIS}{AcadGIS repository}. Full tutorials and worked examples are provided in the \href{https://doc.acadgis.com/tutorials/world-showcase/}{World showcase} and the release notebooks, including \href{https://github.com/riponcm/AcadGIS/blob/v0.2.0/notebooks/All\%20in\%20one.ipynb}{\texttt{All in one.ipynb}}.

Administrative boundaries for Bangladesh, Iraq, India, and the USA are bundled. Other geographic and Earth-observation layers are retrieved from the cited public providers: GADM, Natural Earth, OpenStreetMap, Copernicus GLO-30, ESA WorldCover, and Sentinel-2. All cartographic figures were produced with AcadGIS~0.2.0. Thematic attributes such as income, education, population, and interpolated temperature are synthetically generated in the tutorial notebooks, with fixed random seeds where applicable; they demonstrate expected data structures and do not represent observed measurements. No restricted or confidential research data were used. The Apache-2.0 license applies to the AcadGIS software; upstream geographic data remain subject to their providers' licenses, terms, and attribution requirements.

\section*{Acknowledgements}
AcadGIS builds on the open-source scientific-Python and geospatial ecosystems, and on open geodata from GADM,
Natural Earth, OpenStreetMap, the Copernicus program, and the ESA WorldCover project. During software development, we used ProjectMem as a local-first memory and workflow layer to retain coding decisions, issues, attempted fixes, and project context across sessions \citep{malo2026projectmem}.

\bibliographystyle{unsrtnat}
\bibliography{references}

\end{document}